\documentclass[epj]{svjour}
\usepackage{latexsym}
\usepackage{graphics}
\begin{document}
\title{Mott-insulator transition in Ca$_{2-x}$Sr$_x$RuO$_4$: 
A study of the electronic and magnetic properties}
\author{~V.I.~Anisimov\inst{1} \and ~I.A.~Nekrasov\inst{1} \and 
~D.E.~Kondakov\inst{1}\and T.M.~Rice\inst{2} \and M. Sigrist\inst{2,3}
}                     
%
%
\institute{Institute of Metal Physics, 
Russian Academy of Sciences-Ural Division,
620219 Yekaterinburg GSP-170, Russia \and 
Theoretische Physik , ETH-H\"onggerberg
CH-8093 Z\"urich, Switzerland \and
Yukawa Institute for Theoretical Physics, Kyoto University,
Kyoto 606-8502, Japan}
%
%
\abstract{The electronic structures of the metallic and insulating phases of the
alloy series
Ca$_{2-x}$Sr$_x$RuO$_4$ ($0\leq x \leq 2$) are calculated using LDA,
LDA+U and Dynamical Mean-Field Approximation methods. In the end
members the
groundstate respectively is an orbitally non-degenerate antiferromagnetic
insulator ($x=0$) and a good metal ($x=2$). 
For $x>0.$5 the observed Curie-Weiss paramagnetic
metallic state which possesses a local moment with the unexpected spin
$S=1/2$, is 
explained by the coexistence of localized and itinerant Ru-4d-orbitals.
For $0.2<x<0.5$ we propose a state with partial orbital and
spin ordering. An effective model for the localized orbital and spin
degrees of freedom is discussed. 
The metal-insulator transition at $x=0.2 $ is
attributed to a switch in the orbital occupation associated with a 
structural change of the crystal. 
} 
\maketitle
\section{Introduction}
\label{intro}
The discovery of unconventional superconductivity in \newline Sr$_2$RuO$_4$
\cite{SRO,PhysToday} has evoked considerable interest in the electronic
properties of ruthenates. Curiously the substitution of the smaller
Ca$^{2+}$-ions for Sr$^{2+}$-ions does not lead to a more metallic
state but to an antiferromagnetic (AF) Mott insulator with a staggered
moment of $S=1$ as expected for a localized Ru$^{4+}$-ion which has 4
electrons in the $t_{2g}$-subshell. As will be discussed below, this
insulating behavior is driven by a crystallographic distortion and a
subsequent narrowing of the Ru-4d bands. The complete series of
isoelectronic alloys for intermediate concentrations has recently been
synthesized and studied by Nakatsuji and Maeno~\cite{maeno,CRO1}. 
This gives a rare
opportunity to examine the evolution of the electronic structure
from a multi-band metal to a Mott-insulator transition in an
isoelectronic system.
The evolution is not at all monotonic and the metal-insulator 
transition does not take
place as a simple Mott transition but proceeds through a series of
intermediate regions with unexpected behavior.

The most dramatic example is the system at a concentration $x=x_c=0.5$. At
this critical concentration the susceptibility shows a free Curie form
with a $S=1/2$ moment (not $S=1$) per Ru-ion coexisting with metallic
transport properties. This critical concentration represents the
boundary of the paramagnetic metallic region which evolves as Ca is
substituted in the good metal, Sr$_2$RuO$_4$. At higher Ca
concentrations ($x_c>x>0.2$), the alloys enter a region with AF
correlations at low temperature
but still with metallic properties.  Insulating behavior
appears only at smaller values of $x<0.2$. The challenge that we 
address in this paper, is to understand this unexpected and
nonmonotonic evolution and in particular the exotic behavior in the
vicinity of the critical concentration at $x\approx x_c$.

We begin by discussing the end members. Sr$_2$RuO$_4$ is a good metal.
Its Fermi surface has been determined by de Haas-van Alphen
experiments~\cite{vanalphen}
and agrees very well with the predictions of the local density
approximation (LDA) in the density functional theory~\cite{singh}. 
Ca$_2$RuO$_4$ is
a AF Mott insulator and it can be well described by augmenting the LDA
by a mean field to include the onsite correlation --- the so-called
LDA+U method~\cite{Anisimov91}. 
The third section of this paper is devoted to the examination of
the intermediate concentrations by
performing calculations for a series of characteristic $x$-values. The
most challenging is the region $x\approx x_c=0.5$ where there are
strong correlations but no symmetry breaking so that both LDA and
LDA+U are inapplicable. At this concentration we employ the recently
developed the {\it ab initio} computational scheme combining local density
approximation and dynamical mean field 
theory~\cite{vollhardt,pruschke,georges96} 
(LDA+DMFT)~\cite{LDA+DMFT1,LDA+DMFT2}. We use LDA calculations to
determine the
input parameters in the effective Anderson impurity model which in
turn is treated using a non-crossing approximation 
(NCA)~\cite{prhuschke89}. These
calculational schemes give us reliable information on the evolution of
the electronic distribution among the 3 orbitals in the $t_{2g}$-subshell,
which we shall show is the key to understanding the electronic
properties. 
The paper concludes with a discussion and summary of our
results. A brief account of this work has appeared 
elsewhere~\cite{condmatt}.

\begin{figure}
\resizebox{0.4\textwidth}{!}{%
  \includegraphics{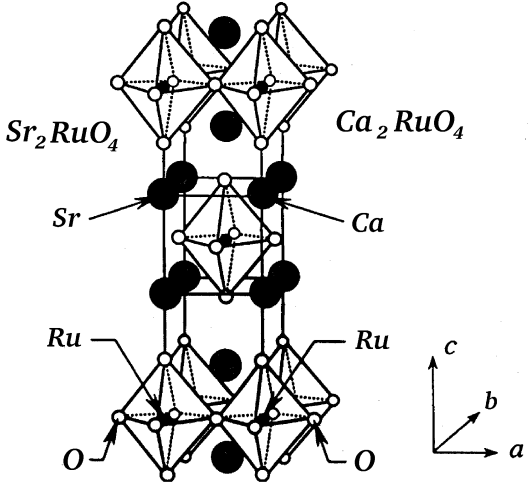}}
\caption{Basic crystall structure of isoelectronic alloy series
         Ca$_{2-x}$Sr$_{x}$RuO$_{4}$.}
\label{structura}
\end{figure}

\section{End Members: Sr$_{\mathbf 2}$RuO$_{\mathbf 4}$ and
  Ca$_{\mathbf 2}$RuO$_{\mathbf 4}$}

\begin{figure}
\resizebox{0.4\textwidth}{!}{%
  \includegraphics{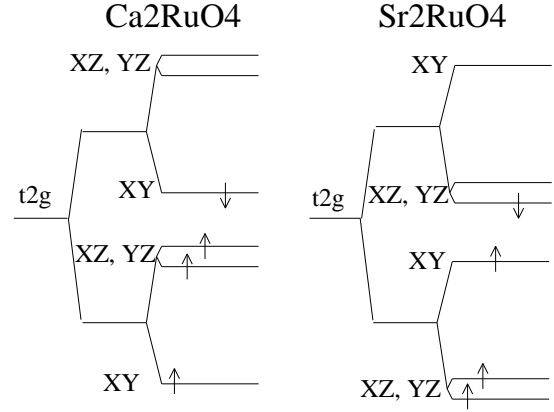}}
\caption{Local electronic structure of isoelectronic alloy series
        Ca$_{2-x}$Sr$_{x}$RuO$_{4}$.
        In Ca$_{2}$RuO$_{4}$ spin-down electron occupies
        $xy$-orbital (left panel);
        In Sr$_{2}$RuO$_{4}$ spin-down electron occupies
        $xz$/$yz$-orbitals (right panel);}
\label{loc_lev_struct}
\end{figure}

\begin{figure}
\resizebox{0.4\textwidth}{!}{%
  \includegraphics{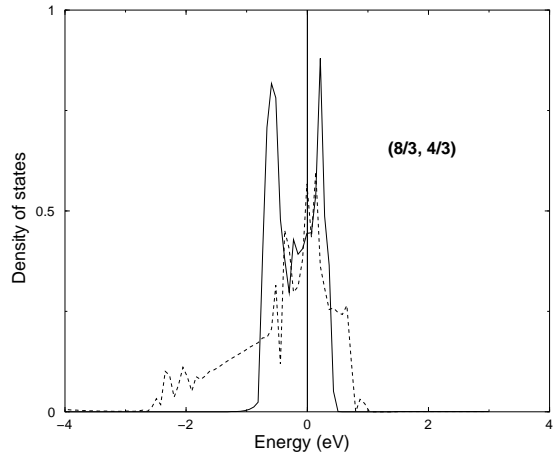}}
\caption{Density of $t_{2g}$ states for Sr$_2$RuO$_4$ obtained from LDA
calculation. The solid line is the DOS for the ($xz, yz$)-orbitals and the
dashed line for the $xy$-orbital.
($n_{(yz,zx)}, n_{xy}$) indicates the electron occupation of the orbitals.}
\label{Sr2RuO4}
\end{figure}

We start with Sr$_2$RuO$_4$ (or $x=2$). This is a good metal, forming
a 3-dimensional but anisotropic Landau-Fermi liquid at low
temperatures~\cite{cor1,cor2}. 
Sr$_2$RuO$_4$ crystallizes in the undistorted
single-layered  K$_2$NiF$_4$-structure~\cite{CRO3,Sr-struc}
(see Fig.~\ref{structura})
with lattice parameters quoted in Table~\ref{cryst}. 
The RuO$_6$-octahedra are
slightly elongated along the $c$-axis. The Ru-ions have a formal
valence Ru$^{4+}$ and have a tetragonal local symmetry. The $2p$-O
levels are completely filled, leaving 4 electrons in $t_{2g}$-subshell
of the 4d-Ru levels. The crystal field level scheme that would apply
for an isolated Ru$^{4+}$-ion is shown in
Fig.~\ref{loc_lev_struct}. 
The upper $e_g$-shell (not included in this figure) is empty. The
splitting between the 
$xy$-orbitals and the degenerate $\{xz,yz\}$-orbitals is small. But
the $xy$-orbitals $\pi$-hybridize with $2p$-orbitals of all 4 in-plane
O-neighbors while the $xz(yz)$-orbitals $\pi$-hybridize only with the
2 O-neighbors along the $x(y)$-axis. As a result the $xy$-bandwidth is
approximately twice the $\{xz,yz\}$ bandwidth (see Fig.~\ref{Sr2RuO4}). 
The LDA calculations \cite{singh}
give 3 Fermi surface sheets, one with essentially $xy$ and two with
mixed $\{ xz, yz\}$ character. Their shape and volume agree with the
de~Haas-van Alphen results~\cite{vanalphen}.

The volumes contained by the Fermi surface sheets give an almost equal
occupancy of each of the 3 $t_{2g}$-orbitals. If we denote the
occupancy of the $\{ xz, yz\}$ and $(xy)$-orbitals by
$(n_{(\alpha,\beta)}, n_\gamma)$, then Sr$_2$RuO$_4$ has the
fractional occupancy $\left( 8/3,\; 4/3 \right)$. Altho' there are
clear signs of strong correlations in the enhanced effective mass
(enhancements $\sim 3-4$~\cite{vanalphen,eff_mass1,eff_mass2}) 
and low effective Fermi temperature, the low-temperature behavior is
clearly that of a well-defined Landau-Fermi liquid.

\begin{figure}
\resizebox{0.4\textwidth}{!}{%
  \includegraphics{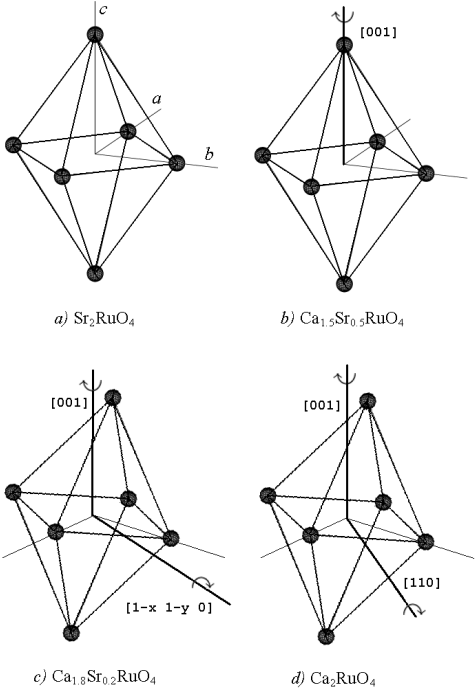}}
\caption{Scheme of crystall distortion of Ca$_{2-x}$Sr$_{x}$RuO$_{4}$
Consecutive structural change of the O-octohedra in the 
alloy series Ca$_{2-x}$Sr$_{x}$RuO$_{4}$:
\textit{a)} Ideal structure K$_{2}$NiF$_{4}$-type (space group I4/$mmm$);
\textit{b)} Space group I4$_{1}$/$acd$ derives from
I4/$mmm$ by rotation around [001]-axis;
\textit{c)} Space group P2$_{1}$/$c$ described by the additional
rotation around a free axis in the octahedron basis plane.
\textit{d)} Space group P$bca$ derived from the ideal structure by rotation
around the [001]- and [110]-axes.}
\label{distorsh}
\end{figure}

Turning to the other end member, Ca$_2$RuO$_4$ or $x=0$, the
substitution of the smaller Ca$^{2+}$-ion for Sr$^{2+}$ does not lead
to a uniform shrinking of the lattice parameter. Instead the
RuO$_6$-octahedra undergo a combined rotation and tilt ($ Pbca
$-structure) so that the 
Ru-O bond length is preserved but the Ru-Ru separation contracts. In
Fig.~\ref{distorsh} 
we illustrate the relevant distortion of the crystal structure.
This distortion bends the Ru-O-Ru bond angle away from 180$^{\rm o}$,
thereby reducing the bandwidth of the $t_{2g}$-orbitals. Also the
smaller size of the Ca$^{2+}$-ion decreases the interlayer distance
(i.e. the $c$-axis lattice constant) which results in a change from
elongation to a compression of the RuO$_6$-octahedra. This in turn
changes the sign of the energy splitting between the $(xy)$- and
$(xz,yz)$-orbitals, so that now the $xy$-orbital lies lower in
energy (see Fig.~\ref{loc_lev_struct}).
The crystal structure is orthorhombic (see Table~\ref{cryst}). All
RuO$_6$-octahedra are equivalent with a rotation around their long
axis $(0\,0\,1)$ and a tilt around the diagonal in-plane axis
$(1\,1\,0)$ (Fig.~\ref{distorsh}d). 
Note all inplane O-ions are equivalent in this structure.

Ca$_2$RuO$_4$ is an AF insulator. The LDA+U method~\cite{Anisimov91}
which is based upon spin-orbital unrestricted Hartree-Fock equation 
(i.e. a static mean field treatment), generally works well for 
magnetic long range ordered insulators~\cite{Anisimov97}. 
We applied this method to Ca$_2$RuO$_4$ choosing parameter
values $U=1.5$ eV and $J=0.7$ eV for the onsite Coulomb repulsion and
the intra-atomic Hund's Rule coupling. The method converged to an AF
insulating groundstate in which the lower $xy$-orbital is fully
occupied and the majority spin $(xz, yz)$-orbitals are also
occupied. The sublattice magnetization is reduced by hybridization
with the O-orbitals from the full value of $2~\mu_B$ expected for
$S=1$, to $1.35~\mu_B$. The energy gap is unusually small (0.17 eV).
Both these values agree well with experiment, $1.3~\mu_B$~\cite{CRO3}
and 0.2~eV~\cite{CRO4} respectively. Further treating spin-orbit 
coupling in
second order perturbation theory gives preferred orientation of the
magnetic moment along the orthorhombic $b$-axis (or $[1\,1\,0]$ in
tetragonal notation) also in agreement with experiment~\cite{Ca_spin}
(see Appendix).

The LDA+U method gives a satisfactory description of the electronic
structure of Ca$_2$RuO$_4$. In terms of our previous notation,
Ca$_2$RuO$_4$ has an integer orbital 
occupancy of (2,2). The key issue that
we address below is the evolution of the electronic structure
between these very different end members.

\begin{figure}
\resizebox{0.4\textwidth}{!}{%
  \includegraphics{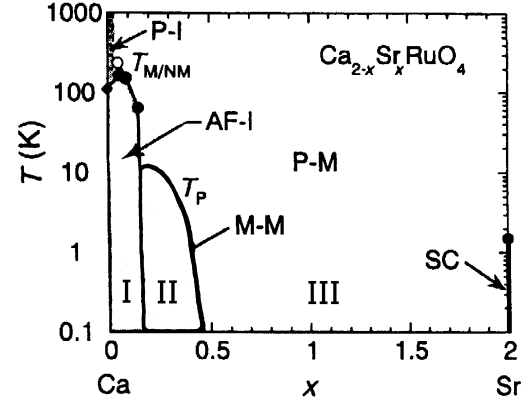}}
\caption{Phase diagram of Ca$_{2-x}$Sr$_{x}$RuO$_{4}$.
The phases are denoted by the abbreviations P-I for paramagnetic
insulating, P-M for paramagnetic metallic, AF-I for antiferromagnetic
insulating, AF-M for antiferromagnetic metallic, M-M for magnetic
metallic and SC for superconducting (Figure is taken from \cite{maeno})}
\label{fig_phase_diag}
\end{figure}

\section{Evolution of the electronic structure at intermediate
  concentrations}

The results of our calculations presented above confirm the
contrasting behavior of the end members, Sr$_2$RuO$_4$ and
Ca$_2$RuO$_4$. 
The experimental investigations of Nakatsuji and Maeno
show a complex evolution of the electronic structure
which they break down into 3
concentration regions, labelled Region I $(0<x<0.2)$, Region II $(0.2
< x<0.5)$ and Region III $(0.5<x<2)$, each with its own characteristic
behavior. Their results are summarized in the phase diagram
(Fig.~\ref{fig_phase_diag}). 
Below we discuss each of these regions, starting with the Sr-rich region.

\subsection{Region III ${\mathbf (2>x>0.5)}$}

The experiments of Nakatsuji and Maeno show that superconductivity is
rapidly suppressed when Ca is substituted for Sr in
Sr$_2$RuO$_4$. This suppression is a natural consequence of disorder for an
unconventional superconductor. More
interesting is the evolution of the spin susceptibility which
progressively increases with Ca substitution and evolves from an
enhanced Pauli susceptibility  into a
Curie-Weiss form. The characteristic (Curie-Weiss) temperature,
$\theta_{cw} (<0) $, 
approaches zero for $x \to 0.5$.  
At the same time
the linear term in the low temperature specific heat is also enhanced
but the Wilson ratio defined at low temperatures is strongly enhanced
and approaches a value of 40 as $x\to x_c$.

This value of $x_c=0.5$ is a critical value, separating a metallic and 
orbitally ordered phase with antiferromagnetic spin correlation $(x<x_c)$
from the paramagnetic metal for 
$x>x_c$. As mentioned above, one observes $\theta_{cw} \approx 0$ for
$x\approx x_c$. Even more remarkable is the evolution of the magnetic
moment 
which takes a value of $S=1/2$ as $x\to x_c$
\cite{maeno_PRB_62,nakatsuji_thesis}. This
value is quite distinct from the value of $S=1$ in Ca$_2$RuO$_4$ ---
the expected value for a localized Ru$^{4+}$-ion with 2 holes in the
$t_{2g}$-subshell.  Moreover, alloys with $x\approx x_c$ are metallic,
not insulating. The explanation of this unexpected behavior presents a
clear challenge.

\begin{table*}
\caption{Crystallographic data, which is used for LDA
calculations: symmetry
group, parameters of lattice, atomic positions and distance between nearest
atoms. Symbol ``/`` denotes that for this structure
corresponding parameter do
not exist.}
\label{cryst}
\begin{tabular}{|c|c|c|c|c|}
\hline
Compound     & $Sr_{2}RuO_{4}$ & $Ca_{1.5}Sr_{0.5}RuO_{4}$ & $Ca_{1.8}Sr_{0.2}RuO_{4}$ & $Ca_{2}RuO_{4}$ \\
\hline
Symmetry group & I4/$mmm$        & $I4_{1}/acd$              & $P2_{1}/c$& $Pbca$          \\
\hline
$a[$\AA$]$        & 3.8603 & 5.3195(1)  & 5.3338(4)  & 5.6323(3)  \\
$b[$\AA$]$        & 3.8603 & 5.3195(1)  & 5.3162(4)  & 11.7463(5) \\
$c[$\AA$]$        & 12.729 & 25.1734(5) & 12.4143(8) & 5.3877(2)  \\
$Vol.[$\AA$^{3}]$ & 189.69 & 712.33(2)  & 352.01(4)  & 356.45     \\
$\beta [^{o}]$    & /      & /          & 90.06(1)   & /          \\
\hline
$Ca(Sr)$ \verb+ + $x$ & 0.0     & 0.0       & 0.0141(21)/0.4903(24) & 0.0593(4) \\
$Ca(Sr)$ \verb+ + $y$ & 0.0     & 0.25      & 0.0137(23)/0.5273(23)& 0.3525(2) \\
$Ca(Sr)$ \verb+ + $z$ & 0.14684 & 0.5492(1) & 0.3483(2)             & 0.0021(5) \\
\hline
$O1$ \verb+  +  $x$ & 0.0    & 0.1933(2) & 0.1939(6)     & 0.3005(4) \\
$O1$ \verb+  +  $y$ & 0.0    & 0.4433(2) & 0.3079(6)     & 0.0272(2) \\
$O1$ \verb+  +  $z$ & 0.3381 & 0.125     & 0.0/0.0196(5) & 0.1952(4) \\
\hline
$O2$ \verb+  +  $x$ & 0.5 & 0         & -0.0344(5) & -0.0212(4) \\
$O2$ \verb+  +  $y$ & 0.0 & 0.25      & -0.0064(7) & 0.1645(2)  \\
$O2$ \verb+  +  $z$ & 0.0 & 0.4568(1) & 0.1649(2)  & -0.0692(3) \\
\hline
$Ru-O1 [$\AA$]$ & 1.930 & 1.929(1) & 1.936(3)/1.926(3) & 2.015(2) \\
                &       &          & 1.941(3)/1.952(3) & 2.018(2) \\
$Ru-O2 [$\AA$]$ & 2.061 & 2.059(3) & 2.056(3)/2.056(3) & 1.972(2) \\
\hline
$Ca-O1 [$\AA$]$ & 2.692 & 2.399(2) & 2.316(7)/2.286(10)  & 2.292(3) \\
                &       & 2.994(2) & 2.445(8)/2.502(9)   & 2.433(3) \\
                &       &          & 2.838(11)/2.934(10) & 2.565(3) \\
                &       &          & 3.141(10)/3.037(10) & 3.313(3) \\
$Ca-O2 [$\AA$]$ & 2.439 & 2.326(4) & 2.294(4)/2.296(4)   & 2.287(3) \\
                & 2.737 & 2.664(1) & 2.416(12)/2.488(13) & 2.362(3) \\
                &       &          & 2.559(13)/2.444(13) & 2.399(3) \\
                &       &          & 2.772(13)/2.845(13) & 3.118(3) \\
                &       &          & 2.932(12)/2.912(13) & 3.296(3) \\
\hline
\end{tabular}
\end{table*}

As discussed above Sr$_2$RuO$_4$ is a good metal with an electronic
structure that divides into two distinct and fractionally filled
bands, a $d_{xy}$-band with $\approx 4/3$ electrons per Ru and the hybridized
$d_{xz,yz}$-band pair with 8/3 electrons per Ru. Now the evolution with Ca
substitution towards a Mott insulating state occurs because the bands
are narrowed by rotation of the RuO$_6$ octahedra. In general, the
evolution from a metal to a Mott-insulator 
is driven by growing Umklapp scattering. This is very clear
in one dimension where studies of chains and ladders show insulating behavior
at half-filling already for arbitrarily small Coulomb repulsion driven
by elastic Umklapp scattering processes across the Fermi surface. Here
we are dealing with an approximately two-dimensional multi-band situation.
In the single band case the Umklapp surface, across which elastic
Umklapp processes are allowed, will in general be different from the
Fermi surface, but at half-filling each encloses the same volume. In
Sr$_2$RuO$_4$ the fractional occupancy of each subband clearly forbids
elastic Umklapp scattering across the Fermi surface in low orders.
Indeed this may well be the reason that Sr$_2$RuO$_4$ forms a good
Landau-Fermi liquid, even though it is clearly close to a Mott
insulator. One way to enhance Umklapp scattering is to transfer
electrons between the subbands. This will allow low-order Umklapp
scattering, when integer occupancy of the subbands is reached.

The theoretical investigation at these concentrations is inhibited by
the lack of symmetry breaking at $x \approx x_c$. Therefore we cannot
use the LDA+U method to build in the onsite correlations that are
totally neglected in standard LDA calculations. Recently considerable
progress has been achieved on the theory of the 
Mott transition in the Hubbard model
by the use of the dynamic mean field theory (DMFT) 
method~\cite{pruschke,georges96}. This is essentially an
expansion around an infinite coordination number and formulates the
problem in terms of an effective Anderson impurity model which is to
be solved self-consistently. In this way the growth of onsite
correlations can be treated as the Mott transition is approached in a
paramagnetic metal. Recent advances use LDA calculations to determine
the input parameters and a non-crossing approximation (NCA) to solve
the effective Anderson model.

\begin{figure}
\resizebox{0.4\textwidth}{!}{%
  \includegraphics{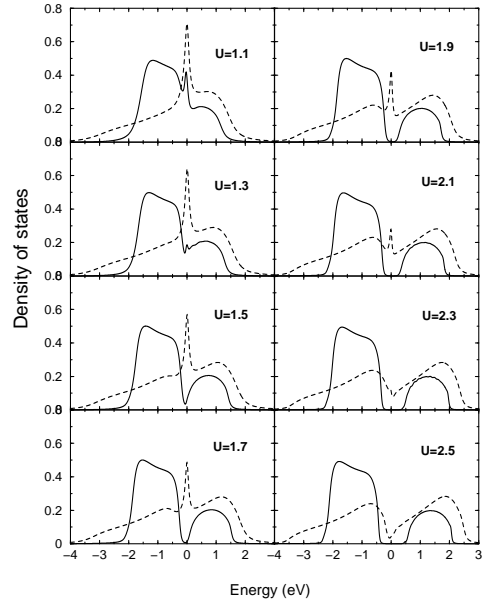}}
\caption{Results of LDA+DMFT(NCA) calculations obtained within
LDA DOS for Sr$_2$RuO$_4$. The solid line is the DOS for $xz, yz$-orbitals and
the dashed line for ($xy$)-orbital.
At $U$ = 1.5 eV the $xz, yz$-orbitals become localized.
At $U$ = 2.5 eV additionally the localization of $xy$-orbital occurs.
The Fermi energy is defined to be zero and was adjusted to
conserve the number of particles (4 electrons per site).}
\label{Sr_DMFT}
\end{figure}

We performed a series of calculations using this LDA + DMFT (NCA)
approximation scheme~\cite{zoelfl99,ldadmftnca}
for the Sr$_2$RuO$_4$ structure. We increased the
value of Hubbard-$U$ to examine how the onsite correlations grow. In
Fig.~\ref{Sr_DMFT} 
is a series of results for the density of states (DOS) in the
$xy$- and $(xz, yz)$-subbands. Since these subbands have quite
different widths, the onset of Mott localization occurs at different
critical values of $U$. Thus we see that as $U$ is increased through a
value of $U\approx 1.5$ eV there is a transfer of electrons between
the subbands so that the 
integer occupancy of 3 electrons and Mott
localization appears in $(xz,yz)$-subbands while the broader
half-filled 
$xy$-band remains itinerant. This unusual behavior is driven by the
combination of the crystal field splitting, as shown in 
Fig.~\ref{loc_lev_struct}
[$(xz,yz)$ lower] and the narrower bandwidth of the
$(xz,yz)$-orbitals. A further increase in the value of $U$ to $U
\approx 2.5$ eV is required to obtain Mott localization also in the
$xy$-subband.

These results lead us naturally to the following proposal to explain
the anomalous properties in the critical concentrations $x=x_c$. The
electronic configuration is now (3,1). The 3 electrons in the
$\{xz,yz\}$-subbands are Mott localized and have a local moment of
$S=1/2$. The remaining valence electrons are in the itinerant 
$xy$-band and is
responsible for the metallic character. Thus at this concentration we
have the unusual situation of localization in only part of the
4d-orbitals and coexisting localized and itinerant 4d-orbitals. Note
that in the orthorhombic crystal structure at $x=x_c$ the 2 subbands
have different parity under reflection around a RuO$_2$-plane, similar
to tetragonal Sr$_2$RuO$_4$, which forbids direct hybridization
between the subbands. This proposal explains in a natural way the
unexpected moment of $S=1/2$ of the Ru-ions and the coexistence of
metallic behavior and local moments. 

Note that the calculations are
carried out more conveniently by increasing the value of the onsite
repulsion, $U$ which however should not change appreciably with the
concentration, $x$. In reality it is the bandwidth which is changing
with the decreasing $x$ as the RuO$_6$-octahedra progressively rotate
when Ca is substituted for Sr. The key result however is the existence
of a parameter range where this partial localization is stable. 
The fact that we calculated only for the highly symmetric
Sr$_2$RuO$_4$ structure, rather than the distorted structure is, we
believe, unimportant in establishing this (3,1) configuration as a
stable electronic configuration.

\subsection{Region II ${\mathbf (0.5>x>0.2)}$}

  At lower values of $x$ we enter Region II $(0.5>x>0.2)$
  characterized by a tilting plus rotation of RuO$_6$-octahedra.
  Ca$_{1.8}$Sr$_{0.2}$RuO$_4$ has a low-symmetry crystal structure
  with the space group $P2_1/c$ \cite{Ca-struc}, which can be obtained
  from the tetragonal $I4/mmm$ structure by rotating and tilting of
  the RuO$_6$-octahedra similar to pure Ca$_2$RuO$_4$ but with a
  smaller tilting angle \cite{Ca-struc} (Fig.~\ref{distorsh}c).  
  There are now two types of
  in-plane oxygen ions and two types inequivalent of
  RuO$_6$-octahedra.  The RuO$_6$-octahedra continue to be elongated
  in this region so that the $xy$-orbital continues to lie higher in
  energy. The metallic character of the alloys in this region shows
  that the itinerant character of the $xy$-subband is preserved,
  although the bandwidth will be narrowed by the additional tilting
  distortion of the RuO$_6$-octahedra. Our conclusion is that the
  (3,1) orbital occupation continues to hold also in Region II with
  localization of the electrons only in the $\{ xz, yz\}$-subband.

\subsection{Region I ~${\mathbf (0.2>x>0)}$~ Ca-rich}

The Ca-rich region is characterized by a transition to an insulating
groundstate and simultaneously a change in the crystal structure. The
S-Pbca structure of the groundstate in this region evolves
continuously out the groundstate of pure Ca$_2$RuO$_4$. As discussed
above, the change from elongated to compressed RuO$_6$-octahedra causes a
switch in LDA+U in the orbital occupation numbers to (2,2) i.e. to a
filled $xy$-subband and a half-filled $\{ xz, yz\}$-complex. Therefore
we assign the insulating groundstate in all of Region I, to an
orbital occupation (2,2). The first order transition between
Regions II and I as $x$ is decreased below 0.2, is to be associated
with a switch in the
orbital occupation from (3,1) to (2,2).

\section{Magnetic Properties of the different regions}

\subsection{Region III}

We now turn to the magnetic properties at $x\geq x_c = 0.5 $. These are
dominated by the local $S=1/2$ moments.  
A Kondo-type of interaction between the two bands $(xz,yz)$ and $xy$
can be excluded due to the absence of hybridization. The $xy$-orbital
cannot hybridize with the other two, since they have opposite parity
with respect to reflection on the basal plane. 
Via Hund's rule coupling, however an 
RKKY interaction between the localized spins is induced, 

\begin{equation}
{\cal H}_{{\rm RKKY}} = - \sum_{{\bf q}} J^2_H \chi({\bf q}) {{\bf
    S}}_{\bf q} \cdot {{\bf S}}_{-{\bf q}}
\end{equation}
where $ {\bf S}_{\bf q} = {\bf S}_{yz, {\bf q}} + {\bf S}_{zx, {\bf
    q}} $ and $ \chi({\bf q}) $ is the static spin susceptibility in the $
d_{xy} $-band and $ J_H $ is the onsite Hund's rule coupling, and
promotes antiferromagnetic correlations.     
At the same time, however, Hund's rule
  coupling may cause the ferromagnetic correlations through the double
  exchange mechanism. Thus, the two types of spin interactions
  mediated by the itinerant electrons of the $xy$-band tend to
  compensate each other, such that the net exchange coupling between
  neighboring localized spins occurs mainly through superexchange
  processes in the $(yz,zx)$-band. The highly anisotropic hopping
  matrix elements between these orbitals, however, leads to an
  essential dependence of the superexchange interaction on the orbital
  configuration of the minority spin electron (or single hole) of each
  Ru$^{4+}$-ion in the degenerate $(xz,yz)$-bands.

In order to gain more insight into the possible form of the orbital 
ordering we performed LDA+U calculations for the critical concentration $
  x_c=0.5$.
Ca$_{1.5}$Sr$_{0.5}$RuO$_4$ has the space group $I4_1/acd$
\cite{Ca-struc}, i.e. the RuO$_6$-octahedra are only rotated
around the $c$-axis with no tilting (Fig.~\ref{distorsh}b). 
The RuO$_6$-octahedra remain elongated to the same degree as for
pure Sr$_2$RuO$_4$ such that 
the $xy$-orbital is still higher in energy than $(xz,yz)$-orbitals 
(Fig.~\ref{loc_lev_struct}, right panel). Our calculation suggests
that, under these conditions, correlation yields the orbital
degeneracy in a (3, 1) state. LDA+U calculations
generally overestimate the stability of the Mott insulating state.
In the present case we find for U = 1.5 eV an insulating groundstate with
a charge gap also in the $(xy)$-subband.

Furthermore, the result of the the LDA+U calculation shows that an
orbital ordering of the AFO-type is favored yielding a
ferromagnetic spin exchange.  
The minority-spin electrons occupy alternating $xz$- and
$yz$-orbitals with a slight tilting of the orbital
planes away from the c-axis. This tilting indicates the partial
admixture of $ d_{xy} $-orbitals within the LDA+U approach.

\subsection{Region II}
  Since magnetic properties are intimately connected to the orbital
  order in $\{ xz, yz\}$-subband, it is necessary to first examine the
  effect of the crystalline distortion on the orbital order. To this
  end we performed LDA+U calculation for this tilted structure and
  obtained a rather complicated orbital order. The groundstate is an
  AF insulator.  The minority-spin electrons (1 per Ru-atom) occupy the
  orbitals whose planes are in average directed along the $a$-axis (in
  tetragonal notation (110) direction).  However on every one of the 4
  Ru-atoms in the unit cell those planes are rotated from the $a$-axis
  by +20$^\circ$ and +15$^\circ$ on one layer and by -20$^\circ$ and
  -15$^\circ$ on the   next layer. On the average the orbital
  orientation corresponds to the state $ d_{yz} + d_{zx} $
  Also on one of the 2 Ru-atoms in every layer there is an  
  additional tilting of the orbital plane from the long $c$-axis on
  34$^\circ$ (see Fig.~\ref{orbital_18_02})
  which corresponds to the admixture of the $xy$-orbital component.
  The calculation of the easy axis using the second order perturbation
  theory for spin-orbit coupling gave the direction of the magnetic moment
  as along the $a$-axis (tetragonal $ [1\bar{1}0] $ direction) with a
  28$^\circ$ tilt from the layer plane (see Appendix). Measurements of
  the uniform 
  magnetic susceptibility show a peak in the temperature dependence
  which is most pronounced for the $ [110]$ direction of the magnetic
  field, in agreement with our LDA+U results \cite{maeno}.

\begin{figure}
\caption{Orbital ordering in Ca$_{1.8}$Sr$_{0.2}$RuO$_4$. At the figure are
presented minority spin 
4-$d$ orbitals of Ru-atoms. $z$-axis directs upwards; $x,y$-axes is
parallel to a bases of oxygen octahedra.}
\label{orbItal_18_02}
\end{figure}

\subsection{Region I}

The orbital configuration (2,2) is non-degenerate as Hund's Rule
determines a fully spin polarized $S=1$ ionic groundstate with a
single minority-spin electron in the $xy$-orbitals. The LDA+U method works
well here and gives an AF Mott insulating groundstate.

Throughout Region I, there is a first order insulator to metal
transition as the temperature is raised accompanied by a switch from
compressed to elongated RuO$_6$-octahedra. It is tempting to interpret
this as a first order transition between the orbital occupancies from
(2,2) to (3,1). However a detailed examination of this proposal has not
been made. In this context it would be also interesting to study the
effect of pressure. Since the low-temperature phase of Ca$_2$RuO$_4$ 
has larger volume than the high-temperature phase (see Tab. I),
pressure tends to stabilize the latter. Assuming that indeed the 
(3,1) electronic distribution is realized in this structural state,
then a orbitally ordered phase accompanied by magnetic order based on
the localized spin-1/2 degree of freedom could be realized at low
enough temperature. Since the crystal structure is close to the one
at $ x=0.2 $ (see Tab. I) one might conclude that AF order would
prevail. However, as we will see in the next section,  
the final form of the exchange interactions between the spins is a
subtle issue, in particular, if a staggered orbital component are
included.      

\begin{table*}
\caption{Regions of Ca$_{2-x}$Sr$_x$RuO$_4$ with
orbital occupancy, localized orbital and spin degrees of freedom
and order or (dominant correlations) (AF = spin antiferromagnetic,
FM = spin ferromagnetic, AFO = antiferro-orbital, FO = ferro-orbital).}
\label{results}
\begin{tabular}{|l|cccc|}
\hline
Region & ($ n_{yz,zx} , n_{xy} $) & orbital & spin & order (correlation) \\
\hline
I $ (0 \leq x < 0.2) $ & (2,2) & -&  $ S=1$ & AF \\
II $ (0.2 \leq x < 0.5 ) $& (3,1) & $ (yz,zx) $ & $ S= \frac{1}{2} $ &
FO, (AF) \\
III $(x \to 0.5)$ & $ (3,1) $ & $ (yz,zx) $ &  $ S
\to \frac{1}{2} $ &  (AFO) , (FM) \\
III $ (x=2) $ & $ \left(\frac{8}{3}, \frac{4}{3} \right) $ & - & S=0 & - \\
\hline
\end{tabular}
\end{table*}

\section{Effective Model for the (3,1) electron distribution}

\subsection{Derivation of spin-isospin Hamiltonian}

In this Section we derive the effective Hamiltonian for the localized
spin and orbital degrees of freedom in the electron configuration
$(n_{(\alpha,\beta)}, n_{\gamma} ) = (3,1) $ which covers the boundary
of region II and III and all of region II. The electrons in the
$ d_{yz} $- and $
d_{zx} $-orbital are localized, while the $ \gamma $-band remains
metallic, although it is half-filled. We ignore here this metallic band
and concentrate on the localized degrees of freedom. The single hole
occupying the two localized orbitals has a spin-1/2 and orbital
index. The latter can be described as isospin-1/2 using 
the notation $ | + \rangle $ for the $ d_{yz} $- and $ |
- \rangle $ for the $ d_{zx} $-orbital. The isospin operator 
$ \mbox{\boldmath $ I $} $ acts on these states by $ I^z |\pm
\rangle = \pm \frac{1}{2} |\pm \rangle $ and is a generator of an SU(2)
transformation in orbital space. 

We will concentrate on the strongest interaction between the localized
degrees of freedom. The 
nearest-neighbor hopping due to $ \pi $-hybridization between the
Ru-$d$ and O-$p$-orbitals lead to the formation of two
independent quasi-one-dimensional bands, with dispersion in $y$
($x$)-direction for the $ d_{yz} $ ($ d_{zx}$)-orbital. 
Including the onsite interactions the Hamiltonian for
these two orbitals has the form

\begin{eqnarray}
{\cal H} & = & - t \sum_{i,{\bf a},s} (c^{\dag}_{i+a_x, -,s} c_{i,-,s} + 
c^{\dag}_{i+a_y,+,x, +,s} c_{i,+,s} + h.c.) \nonumber \\
&& + U \sum_{i, \mu=+,-} n_{i,\mu,\uparrow} n_{i,\mu,\downarrow} 
+U' \sum_{i} n_{i,+,s} n_{i,-,s'} \\
&& - 2 J_H \sum_i ({\bf S}_{i,+} \cdot
{\bf S}_{i,-} + \frac{1}{4} n_{i,+} n_{i,-}) \nonumber
\end{eqnarray}
where $ t $ denotes the hopping matrix element, $ U $ and $ U' $
are the onsite intra- and interorbital Coulomb repulsion and $ J_H $
is the Hund's rule coupling (the vector $ {\bf a} = (1,0) $ and $
(0,1) $ connectes the nearest neighbor sites). We can reduce the
number of parameters by the relation $ U = U' + 2 J_H $.  

We now consider the case of a nearest-neighbor bond along the $x$-direction and
calculation the effective spin-spin interaction for different orbital
configurations. For illustration of the processes involved we
introduce the following self-explanatory notation shown in the two examples, 

\begin{eqnarray} 
|+,\uparrow \rangle_i \otimes |+,\downarrow \rangle_{i+a_x} = \left| 
\begin{array}{cc} \uparrow & \downarrow \\ \uparrow \downarrow & 
\uparrow \downarrow \end{array} \right\rangle \\
|+,\uparrow \rangle_i \otimes |-,\uparrow \rangle_{i+a_x} = \left| 
\begin{array}{cc} \uparrow & \uparrow \downarrow \\ \uparrow \downarrow & 
\uparrow \end{array} \right\rangle 
\end{eqnarray}
The upper (lower) row in the isospinor notation correspond to the
$d_{yz} $ ($d_{zx}$)-orbital. Now we discuss the virtual exchange
processes leading to the spin interactions for the $x$-bond.

(1) The configuration $ |+,s \rangle \otimes |+, s' \rangle $ does not
lead to any interaction. (2) The configuration $ |-,s \rangle \otimes 
|- ,s' \rangle $ yields antiferromagnetic superexchange through the
exchange path,

\begin{equation}
\left| \begin{array}{cc} \uparrow \downarrow & \uparrow \downarrow \\
\uparrow & \downarrow \end{array} \right\rangle
\stackrel{-t}{\rightarrow} 
\left| \begin{array}{cc} \uparrow \downarrow & \uparrow \downarrow \\
0 & \uparrow \downarrow \end{array} \right\rangle \\
\stackrel{-t}{\rightarrow} 
\left| \begin{array}{cc} \uparrow \downarrow & \uparrow \downarrow \\
\downarrow & \uparrow \end{array} \right\rangle.
\end{equation}
This leads to an effective Hamiltonian
\begin{equation}
{\cal H}' = J({\bf S}_i \cdot {\bf S}_{i+a_x} - \frac{1}{4} )
\end{equation}
with an exchange coupling constant 
\begin{equation}
J = \frac{4 t^2}{U}.
\end{equation}
(3) Finally the configuration $ |+,s \rangle \otimes |-,s' \rangle $ 
gives an ferromagnetic spin coupling by the following type of path,

\begin{equation}
\left| \begin{array}{cc} \uparrow & \uparrow \downarrow \\
\uparrow \downarrow & \uparrow \end{array} \right\rangle
\stackrel{-t}{\rightarrow} 
\left| \begin{array}{cc} \uparrow & \uparrow \downarrow \\
\uparrow & \uparrow \downarrow \end{array} \right\rangle \\
\stackrel{-t}{\rightarrow} 
\left| \begin{array}{cc} \uparrow & \uparrow \downarrow \\
\uparrow \downarrow & \uparrow \end{array} \right\rangle.
\end{equation}
Here the intermediate state has lower energy, if the two spins are
in a triplet configuration due to the Hund's rule coupling. 
The interaction energies for the intermediate states corresponding
to the spin singlet and triplet state are
\begin{eqnarray}
\Delta E_{\rm singlet} = U' + J_H \\
\Delta E_{\rm triplet} = U' - J_H 
\end{eqnarray}
so that the effective Hamiltonian can be cast into the form
\begin{equation}
{\cal H}'' = J' \left( {\bf S}_i \cdot {\bf S}_{i+a_x} +
  \frac{1+\alpha}{2(1-\alpha)}  \right)
\end{equation}
with 
\begin{equation}
J' = - \frac{2 t^2 J_H}{U'^2 - J_H^2} = - J \frac{1 - \alpha}{(3
  \alpha - 1)(\alpha  + 1)} 
\end{equation}
where $ \alpha = U'/U < 1 $. Note that $ \alpha $ should be sufficiently 
larger than $ 1/3 $ in order that the second-order perturbation
approach is valid and the electron distribution (3,1) is unique. 
The orbital configurations in the basis of $ |+\rangle $ and
$ |- \rangle $ are not affected in this second order perturbation
scheme. Therefore the effective isospin Hamiltonian is Ising-like. 
 
Turning to the bond along the $y$-axis we need only to exchange the
role of the two orbitals ($ d_{yz} $ and $ d_{zx} $) and obtain the
same result. Therefore we 
can write the complete effective Hamiltonian for the localized degrees
of freedom as,

\begin{eqnarray}
{\cal H} & = & J \sum_{i,{\bf a}} \left[ \left\{A (I^z_{i+{\bf a}} +
    \eta_{{\bf a}})(I^z_{i} + 
\eta_{{\bf a}}) +B \right\} 
{\bf S}_{i+{\bf a}} \cdot {\bf S}_{i} \right.  \nonumber \\
& & \left. + [C (I^z_{i+{\bf a}} +
\eta'_{{\bf a}})(I^z_i + \eta'_{{\bf a}}) + D \right]
\label{Heff}
\end{eqnarray}
where the coefficients are given by
\begin{eqnarray}
A &=& \frac{3 \alpha^2 + 1}{(3 \alpha-1)(\alpha +1)}  \label{pm1} \\
B &=& \frac{-(1- \alpha)^2}{(3 \alpha^2 + 1)(3 \alpha-1)(\alpha +1)}
\label{pm2} \\
C &=& \frac{5 - 3\alpha}{4 (3 \alpha - 1)} \label{pm3}  \\
D &=& - \frac{2}{(5 - 3 \alpha)(3 \alpha - 1)} \label{pm4}  \\
\eta_{{\bf a}} &=& - \frac{(3 \alpha - 1)(\alpha + 1)}{2(3 \alpha^2
  +1)} (a_x^2 - a_y^2) \label{pm5} \\
\eta'_{{\bf a}} &=& - \frac{5 - 3\alpha}{3\alpha-1} (a_x^2 - a_y^2)
 \label{pm6} 
\end{eqnarray}
Note that the coefficients $ \eta_{{\bf a}} $ and $ \eta'_{{\bf a}} $
have opposite sign for the $x$- and $y$-axis bonds. 

Lattice distortions breaking the tetragonal symmetry yield a bias for
the local orbital  
configuration. For the two basic orthorhombic distortions,
described by the lattice strain combinations $ \epsilon_1 = \epsilon_{xx} -
\epsilon_{yy} $ and $ \epsilon_2 = \epsilon_{xy} $ (tetragonal
notation), we can add the following 
coupling terms to the Hamiltonian, 

\begin{equation}
{\cal H}_{dist} = \sum_i [K_1 \epsilon_1 I^z_i 
+ K_2 \epsilon_2 I^x_i ]
\label{dist}
\end{equation}
where $ K_1 $ and $ K_2 $ are coupling constants. Note that the first
term corresponds to a uniform field parallel to the $z$-axis of the
isospin, while the second is a transverse field. Both drive a
ferro-orbital correlation.

\subsection{Mean field discussion}

We analyze now the properties of our effective model by means of
mean field theory ignoring fluctuations and the influence of
the itinerant $xy$-band. This discussion shows that the main features
of the different phases where the electron distribution (3,1) is
applicable are indeed described well within our reduced effective
model. A discussion beyond the mean field level will be given
elsewhere.  

\subsubsection{Tetragonal system}

\begin{figure}
\resizebox{0.4\textwidth}{!}{%
  \includegraphics{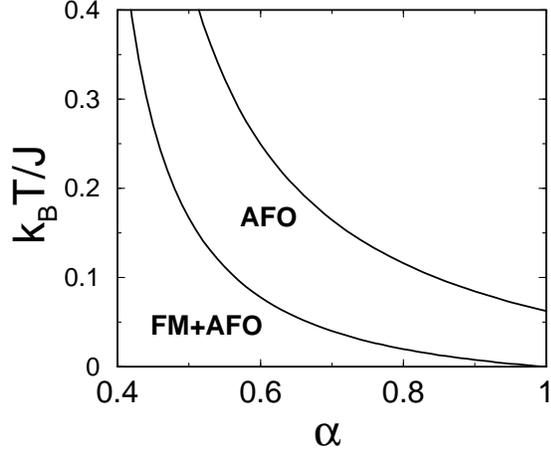}}
\caption{Mean field phase diagram temperature versus $ \alpha=U'/U $
  for $x=0.5$.}
\label{pdtetra}
\end{figure}

The analysis of the parameters of the effective Hamiltonian shows that
a higher energy scale is associated with the isospins than with the
spins. Their
basic interaction is of Ising-AFO-type, because $ C > 0 $
for all $ \alpha $ in the proper range. As a consequence the spins 
correlate ferromagnetically at a lower energy scale. Hence, we
will decouple the effective Hamiltonian for the spin and isospin with
mean values, 

\begin{equation}
m_I = \langle I^z_i \rangle \quad \mbox{and} \quad s = \langle S^z_i
\rangle 
\end{equation}
where $ m_I $ is staggered with opposite sign on the two sublattices
and $ s$ is uniform. The coupled self-consistent equations are
obtained readily, 

\begin{eqnarray}
m_I = \frac{ \sum_{r=\pm 1} r e^{\beta  C r m_I} \cosh \left(\beta
            \left(A \left(\eta_{{\bf 
            a}}^2 - \frac{r m_I}{2}\right)+B \right)s \right)}{2
\sum_{r=\pm 1} e^{\beta  C r m_I} \cosh \left(\beta \left(A \left(\eta_{{\bf
            a}}^2 - \frac{r m_I}{2}\right)+B \right)s \right)} \\
s= \frac{ \sum_{r = \pm 1} e^{\beta C r m_I} \sinh \left(\beta \left(A \left(\eta_{{\bf
            a}}^2 - \frac{r m_I}{2}\right)+B \right)s \right)}{2
\sum_{r = \pm 1} e^{\beta C r m_I} \cosh \left(\beta \left(A \left(\eta_{{\bf
            a}}^2 - \frac{r m_I}{2}\right)+B \right)s \right)}
\end{eqnarray}
where $ \beta = J/k_B T $ is the inverse temperature in units of 
the exchange energy $ J $. The isospin order has the higher transition
temperature $ T_o $ determined by the equation

\begin{equation}
k_B T_o = \frac{CJ}{4}.
\end{equation}
The onset of FM order occurs at much lower
temperature so that it is justified to consider $ m_I $ as already
saturated, $ m_I = 1/2 $. Then
Curie-temperature for the spin is given by

\begin{equation}
k_B T_{\rm C} = - \frac{J}{4} \left[ a \left( \eta_{{\bf a}}^2 -
  \frac{1}{4} \right) + B \right] .
\end{equation}
In Fig. \ref{pdtetra} we show the phase diagram of temperature versus 
the parameter $ \alpha = U'/U $ which determines the coupling
constants. The tendency towards FM order is in good agreement with the 
LDA+U result for $ x = 0.5 $, apart from the fact that the
contributions of $ xy $-orbital have been ignored. 
This analysis suggests that
the orbital order is governed by the exchange processes included in
our Hamiltonian. 

\subsubsection{Crystal distortion}

Let us now consider the influence of a spontaneous crystal deformation
which characterizes region II ($0.2 < x < 0.5 $) 
We will not discuss the origin of the distortion, but take it as given
with a strength which depends on $x$. The FO bias introduced by a
uniform lattice 
distortion is in competition with the AFO exchange interaction. In particular,
a finite strain $ \epsilon_2 $ would introduce a transverse field in
the effective Ising Hamiltonian for orbital interactions. 
An Ising-model in a transverse field has a quantum critical point
beyond which the AFO order would be suppressed for any temperature. 

We consider here the general case where both $ \epsilon_1 $ and $
\epsilon_2 $ are finite, as realized in Region II. 
We distinguish now between the two sublattices $A$ and $B$ on the square
lattice and assign different mean fields to them. Analogous to the case above
we may separate the spin and the isospin problem. First we consider
the isospin problem in the absence of spin correlation. The
corresponding self-consistent equations have the form,

\begin{equation}
m_{IA} = f(\beta, m_{IB}) \quad \mbox{and} \quad m_{IB} = f(\beta,
m_{IA})
\label{FO1}
\end{equation}
with

\begin{equation} \begin{array}{ll} 
f(\beta,m)= & \displaystyle - \frac{2Cm+K_1 \epsilon_1}{2
    \sqrt{(2 C m + K_1    
    \epsilon_1 )^2 + K_2^2 \epsilon_2^2 }} \\ & \\
& \displaystyle \times 
    \tanh\left(\beta \sqrt{(2 C m + K_1
    \epsilon_1)^2 + K_2^2 \epsilon_2^2} \right).
\end{array}
\label{FO2}
\end{equation}
It is easy to verify from this set of equations 
that for finite crystal distortion there is
a uniform component $ m = (m_{IA} + m_{IB})/2 $ at all temperatures and
that the staggered moment $ m_I = (m_{IA}-m_{IB})/2 $ occurs only at
low temperature or not at all, depending on which side of the quantum
critical point the system lies. Hence the staggered moment may
play a minor role on the background of the uniform FO orbital
arrangement. 

\begin{figure}
\resizebox{0.4\textwidth}{!}{%
  \includegraphics{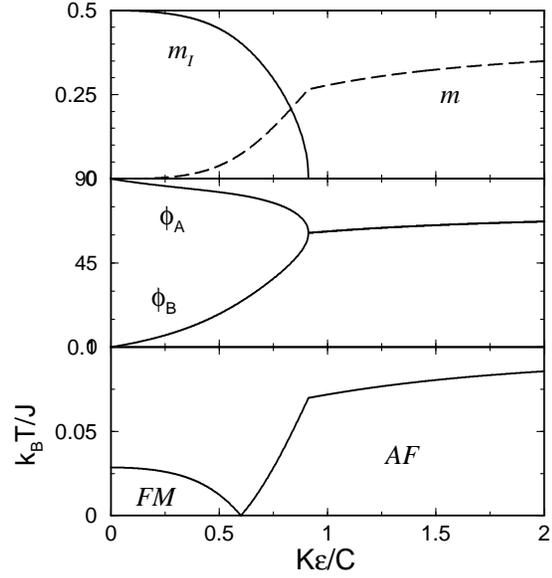}}
\caption{Effect of lattice deformation $ K_1 \epsilon_1 = 1.5 K_2
  \epsilon_2 = K \epsilon $ on the orbital order and the effective
  spin exchange. a) Staggered orbital moment $ m_I $ and uniform
  orbital moment $ m $; b) Orientation of the orbital plane $ \cos
  \phi |+ \rangle + \sin \phi | - \rangle $ for the sublattices $
  A $ and $ B $; c) effective spin exchange ferromagnetic and
  antiferromagnetic}
\end{figure}

The numerical solution of Eq.(\ref{FO1},\ref{FO2}) is shown in
Fig.9. assuming that both $ \epsilon_1 $ and $ \epsilon_2 $ are
finite with a ratio $ K_1 \epsilon_1 / K_2 \epsilon_2 = 3/2 $.
Turning $ \epsilon_{1,2} $ gradually on we observe that the
staggered AFO component is diminished in favor of the FO arrangement
(Fig.9a) with a critical point where the staggered component
completely vanishes. Fig.9b shows the orbital orientation on the
two sublattices $ |d \rangle_{A,B} $ $= \cos \phi_{A,B} | + \rangle + 
\sin \phi_{A,B} | - \rangle $. Obviously, very close to the quantum
critical point we find a dominant FO arrangement with a weak staggered
component. This describes qualitatively the favored ordering pattern
in the LDA+U calculation for $ x=0.2 $, where in each layer 
the orbital plane has an alternating orientation of $ 20^{\circ} $ and $
15^{\circ} $ with respect to the [110] (FO) direction. The comparison
with our mean field data suggests that the system is then rather close
to the quantum critical point. 

Assuming again that at the energy scale where spin correlation sets in
the orbital pattern is basically established, we draw the diagram of transition
temperature versus the strain in Fig.9c. Then we analyze the spin
correlation fixing the orbital order by the $ T=0 $-solution of
Eq.(\ref{FO1},\ref{FO2}). This is again justified by the observation that
the isospin mean field has saturated at the temperature when spin
order sets in.  We find weak 
ferromagnetic coupling for small distortion and stronger antiferromagnetic
coupling for larger distortion. The crossover in the spin interaction 
is still in the
region of rather strong staggered orbital correlations. 
Note that the crossover point between FM and AF spin exchange depends
on $ \alpha $ and move towards larger $ K \epsilon $ for smaller $
\alpha $.  

Guided by this phase diagram we may interpret the region II in terms
of increasing strain as we progress from x=0.5 to 0.2. At x=0.2 the quantum
critical point has been approached but not passed. This interpretation is 
entirely based on the localized orbitals of the $ \alpha
$-$\beta$-bands. From the LDA+U calculation we conclude, however, that
the $ \gamma 
$ -band is also involved in the spin correlations, as is seen in the
tilting of the orbital plane away from the $z$-axis for the minority
spins. As mentioned earlier the $ \gamma$-band should introduce an
RKKY-interaction via the Hund's rule coupling, which is most likely 
antiferromagnetic, thus further diminishing the FM correlation. 

An interesting feature occurs in our effective model in connection
with the sign change of the spin exchange interaction as mentioned
above. Ca$_{2-x}$Sr$_x$RuO$_4$ as a random alloy does not provide 
a uniform crystal deformation, 
but the strain depends on the local Sr and Ca distribution around each
Ru-ion. 
Thus, in the regime of switching between FM and AF coupling, this
can lead to frustration effects at low temperature. Recent
experimental data indicate indeed a glass-like behavior \cite{GLASS}.

\section{Discussion}

We have presented a consistent picture for the unusual phases of the
isoelectronic alloy series Ca$_{2-x}$Sr$_2$RuO$_4$ based on the full
multi-band electronic structure (see Table~\ref{results}). 
Starting from the good
metal Sr$_2$RuO$_4$, we find the effect of Ca-substitution is to
transfer electrons from the wider $xy$-band to the narrower
$(xz,yz)$ bands until at a critical value of $x_c=0.5$ there is
integer occupancy of both subbands. The progressive rotation of the
RuO$_6$ octahedra in this region leads to Mott localization of the 3
electrons in the narrower $(xz,yz)$ bands while the wider $xy$-band
which is now half-filled, remains metallic. This partial localization
of the 4d electrons can explain the puzzling observation of the
coexistence of free $S=1/2$ local moments and metallic behavior in
Ca$_{1.5}$Sr$_{0.5}$RuO$_4$.  The actual metal to insulator transition
occurs between the Ca-rich Regions I ($ 0 \leq x < 0.2 $) and II ($
0.2 \leq x <0.5 $).  We interpret this transition as driven by a
change in the orbital occupancy to a completely filled $xy$-band
leaving 2 electrons in the localized $(xz, yz)$ bands. In this case
the ordered local spin has the conventional value,
$S=1$, whereas in the metallic Region II
short-range correlations of a local $S=1/2$ spin combined with orbital
order are realized, generating a pronounced anisotropy in the magnetic
response.

One of the curious aspects of Region II is the fact that the
half-filled $ xy $-orbital remains itinerant. It is not easy to
test experimentally which among the bands is really responsible for
the metallic behavior. The basic feature of the state we propose is
the presence of a single Fermi surface of electron-like character in
the $xy $- or $ \gamma $-sheet. 
One method which might be able to probe this proposal is
angle-resolved photo emission spectroscopy (ARPES).   

Recently the electronic structure and magnetic properties of
Ca$_{2-x}$Sr$_2$RuO$_4$ have also been investigated by LSDA calculations
\cite{terakura}. The authors have concluded that the crystal structure 
distortions (rotation, tilt and compression
of the O-octahedra) when Ca is substituted for Sr lead to a narrowing
and shifting of the bands and hence to the enhancement of ferro-
and antiferromagnetic instabilities. However these calculations 
do not include the effects of onsite Coulomb interactions between
d-electrons which are responsible for the suppression of charge
fluctuations and for the Mott metal-insulator transition
as Ca$_2$RuO$_4$ is approached.

We would like to thank Y. Maeno, S. Nakatsuji, T. Nomura, M. Braden, 
D. Vollhardt, T. Pruschke, M. Z\"olfl, J. Schmalian 
and A. Poteryaev for many helpful discussions.  This work was
financially supported by the Russian Foundation for Basis Research
grant RFFI-01-02-17063, 
a Grant-in-Aid of the Ministry of
Education, Science, Sports and Culture of Japan (No. 12046238), Netherland 
Organization for the Advance of Pure Science (NWO 047-008-012), 
the Sonderforschungsbereich 484 of the Deutsche Forschungsgemeinschaft and
the Swiss Nationalfonds.

\appendix

\section{Effect of spin-orbit coupling}

For the Ru $4d$-$t_{2g}$-orbitals spin-orbit coupling plays an
important role. If we restrict to the three $ t_{2g} $-orbitals, the
onsite spin-orbit coupling Hamiltonian can is given by

\begin{equation}
{\cal H}_{s-o} = \lambda \sum_i \sum_{a,b,c} \sum_{s,s'} \epsilon_{abc}
c^{\dag}_{ia,s} \sigma^c_{ss'} c_{ib,s'}
\end{equation}
where $ \lambda $ is the coupling constant and $ \epsilon_{abc} $ 
the completely antisymmetric tensor ($a$, $b$, $c$ $ = \{x,y,z \}$).
We identify the indices $x$, $y$ and $z$ with the orbitals $ d_{yz} $, 
$ d_{zx} $ and $ d_{xy} $, respectively. 

In the framework of LDA+U calculation the easy axis direction
was calculated via minimizing the energy of spin-orbit coupling
in the second order perturbation theory as a function of Euler
angles $(\alpha ,\beta ,\gamma)$ defining the direction of magnetization:

\begin{eqnarray}
E(\alpha ,\beta ,\gamma)=
\sum_{\vec k}\sum_{n \sigma , n' \sigma '}\frac{f_{n \sigma \vec k}
- f_{n' \sigma ' \vec k}}{E_{n \sigma}(\vec k) - E_{n' \sigma '}(\vec k)} 
\nonumber \\
|\langle \Psi_{n \sigma}^{\vec k}|{\mathbf{\hat{l}}} \cdot {\mathbf{\hat{s'}}}
| \Psi_{n' \sigma '}^{\vec k} \rangle|^2, \\
 \nonumber \\
{\mathbf{\hat{l}}}\cdot{\mathbf{\hat{s'}}}=\sum_{\mu}\hat{l}_{\mu}
\cdot\hat{s'}_{\mu}=
\sum_{\mu \nu}\hat{l}_{\mu}\cdot
U_{\mu\nu}(\alpha ,\beta ,\gamma)\hat{s}_{\nu}
\end{eqnarray}

where $E_{n \sigma}(\vec k)$ is the energy of $n$-th band for spin projection
$\sigma$ at point $\vec k$ in the Brillouin zone, $f_{n \sigma \vec k}$ is
equal 1 if the corresponding band is occupied (its energy is below the Fermi
level) and is equal 0 in the opposite case. $\Psi_{n' \sigma '}^{\vec k}$ is
the Bloch wave function, $\hat{l}_{\mu}$ and $\hat{s}_{\nu}$ are components
of vector operators of orbital and spin moments, $U_{\mu\nu}(\alpha ,\beta ,\gamma)$
is transformation matrix for rotation of a vector for Euler
angles $(\alpha ,\beta ,\gamma)$.

%
%

\end{document}